\documentclass[]{spie}  

\usepackage{amsmath,amsfonts,amssymb}
\usepackage{graphicx}
\usepackage[colorlinks=true, allcolors=blue]{hyperref}
\usepackage{subfig}
\usepackage{aas_macros}
\usepackage{textcomp}

\title{	In search of the third dimension in Compton X-ray polarimeters}

\author[a]{Fabio~Muleri}
\author[b]{Riccardo~Campana}
\author[b]{Edoardo~Borciani}
\author[b]{Lorenzo~Amati}
\author[a]{Enrico~Costa}
\author[a]{Nicolas~De~Angelis}
\author[a]{Ettore~Del~Monte}
\author[a]{Alessandro~Di~Marco}
\author[a]{Sergio~Fabiani}
\author[c]{Mauro~Fiorini}
\author[a,e,f]{Dawoon~E.~Kim}
\author[a]{Riccardo~Ferrazzoli}
\author[a]{Fabio~La~Monaca}
\author[a]{Carlo~Lefevre}
\author[a,g]{Giovanni~Lombardi}
\author[a]{Hemanth~Manikantan}
\author[b]{Ezequiel~J.~Marchesini}
\author[a]{Romana~Mikusincova}
\author[a]{Alfredo~Morbidini}
\author[a]{Alessio~Nuti}
\author[a,d]{John~Rankin}
\author[a]{Ajay~Ratheesh}
\author[a]{Alda~Rubini}
\author[a]{Paolo~Soffitta}
\author[b]{Enrico~Virgilli}

\affil[a]{INAF - Istituto di Astrofisica e Planetologia Spaziali, via Fosso del Cavaliere 100, Roma (Italy)}
\affil[b]{INAF - Osservatorio di Astrofisica e Scienza dello Spazio,  via P. Gobetti 101, Bologna (Italy)}
\affil[c]{INAF - Istituto di Astrofisica Spaziale e Fisica Cosmica Milano, Via A. Corti 12, Milano (Italy)}
\affil[d]{INAF - Osservatorio Astronomico di Brera, via E. Bianchi 46, Merate (Italy)}
\affil[e]{Dipartimento di Fisica, Università degli Studi di Roma “La Sapienza”, Piazzale A. Moro 5, Roma, (Italy)}
\affil[f]{Dipartimento di Fisica, Università degli Studi di Roma “Tor Vergata”, Via della Ricerca Scientifica 1, Roma, (Italy)}
\affil[g]{Dipartimento di Ingegneria dell'Impresa Mario Lucenti, Università degli Studi di Roma ``Tor
Vergata'', Via Cracovia 50, 00133 Roma, Italy}

\newcommand{\B}{\text{bot}}
\newcommand{\T}{\text{top}}

\authorinfo{Further author information: (Send correspondence to Fabio Muleri)\\Fabio Muleri: E-mail: \url{fabio.muleri@inaf.it}, Telephone: +39 06 4993 4565}

\begin{document} 
\maketitle

\begin{abstract}
Compton polarimeters are typically designed to be sensitive only to the azimuthal angle of the scattered photon, ignoring the scattering angle. Such a 2-dimensional reconstruction of the event is pursued for both simplicity and because the polarization of the incident photon influences only the azimuthal response of the instrument. While this is true for on-axis sources, when the source starts to be off-axis of several degrees the azimuthal response of the instrument is effectively a convolution of the azimuthal and polar scattering angles: measuring the latter would provide a better sensitivity and smaller systematic effects. In this contribution, we will present a design which allows to estimate the scattering angle in a Compton polarimeter through the read-out of the light signal at the two ends of scintillator bars. Such a design is being tested with a representative set-up and first results on the performance are presented.
\end{abstract}


\section{INTRODUCTION} \label{sec:intro}  

The measurement of the polarization of the emission from astronomical sources  at high energy has known a rapid development in the last years. The launch of the NASA-ASI Imaging X-ray Polarimetry Explorer mission (IXPE\cite{Weisskopf2022}) in December 2021 finally has paved the way to measurements in the 2--8 keV energy range for a number of sources belonging to essentially all the classes of X-ray cosmic sources. The outstanding results of this mission on both galactic\cite{Krawczynski2022, Taverna2022, Xie2022} and extra-galactic\cite{Liodakis2022, Marin2023} sources confirmed the long expectation that a significant polarization was to be found in the majority of high energy cosmic sources, and its detection would have provide us unique insights in the emission mechanism and geometry of these sources. With a planned launch in 2027, the Compton Spectrometer and Imager (COSI\cite{Tomsick2023}) is another small-explorer mission bound to carry out imaging, spectroscopy and polarimetry of astrophysical sources in the energy band 0.2--5~MeV. Beyond these (relatively) large missions, a number of instruments have been designed or adapted as a secondary priority to measure polarization. They either targeted specific cosmic sources on-board, e.g., scientific balloons or CubeSats, or bright transient sources like Gamma Ray Bursts (GRBs). Just to name a few of them, there are PolarLight \cite{Feng2019}, PoGO+\cite{Friis2018}, POLAR/POLAR-2\cite{Kole2020, DeAngelis2021}, Astrosat\cite{Vadawale2015} or X-Calibur \cite{Abarr2021}. 

With the exception of instruments primary designed as Compton telescopes, all instruments dedicated to the detection of polarization in the high-energy range have been historically focused only on the detection of the azimuthal response of the instrument. In fact, the detection of the polar distribution of the events was considered an unnecessary complication, as the azimuthal one is the only sensitive to the polarization of the interacting photons. However, it has been pointed out that this is true only in case of photons which are incident orthogonal to the instrument\cite{Muleri2014}, and therefore ``classical'' designs are subject to systematic effects when the source is off-axis of several degrees or more, as it occurs for instruments with a large field of view dedicated to transient sources\cite{Yonetoku2006, Xiong2009}. 

\begin{figure} [ht]
   \begin{center}
    \subfloat[\label{fig:compton_pol}]{\includegraphics[height=6cm]{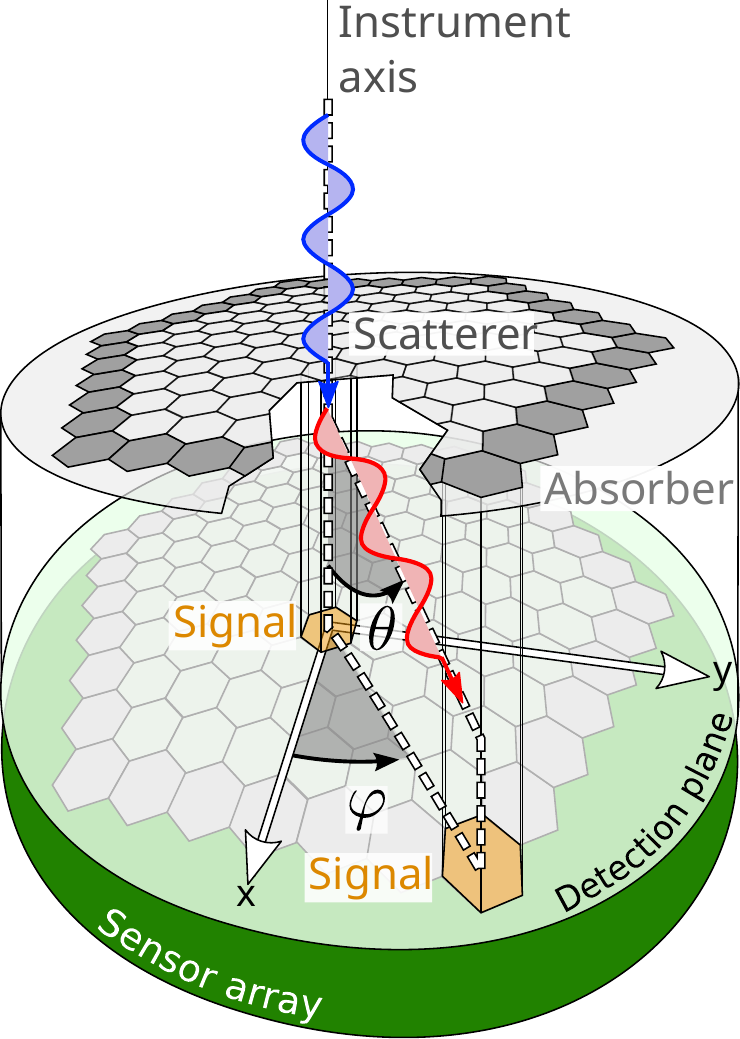}} \hspace{2mm}
    \subfloat[\label{fig:current_vs_proposed}]{\includegraphics[height=6cm]{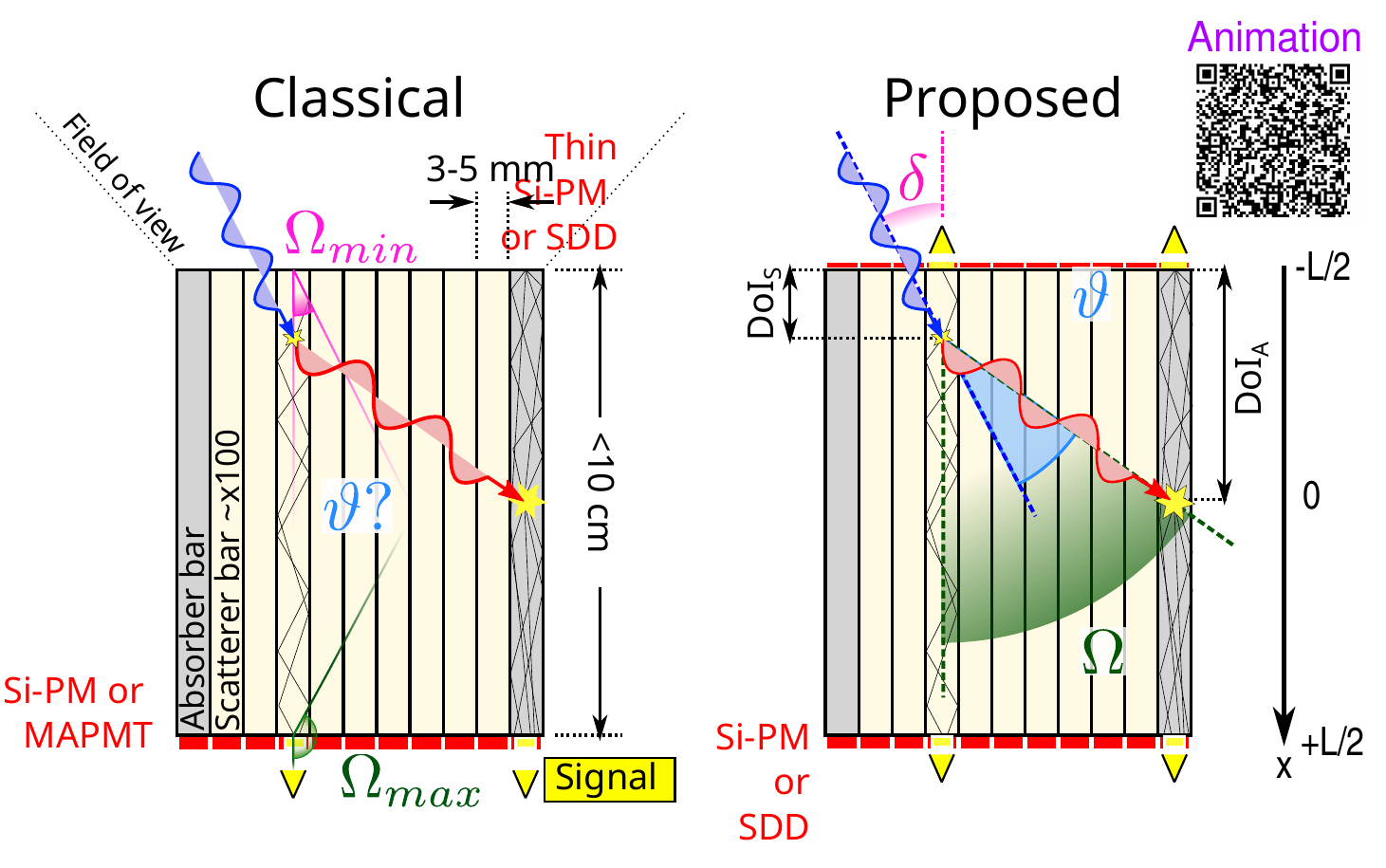}}
	\end{center}
   \caption{(a) A possible implementation of a ``classical'' design for Compton polarimeters working in the hard X-rays. (b) Comparison of the ``classical'' design and the one proposed and under investigation.}
   \end{figure}

A possible implementation of a ``classical'' design for Compton polarimeters working in the hard X-rays is shown in Figure~\ref{fig:compton_pol}. The polarimeter
comprises of an array of bars made of some scintillator material arranged parallel to the instrument axis and read-out from the bottom; inner bars are used to Compton-scatter the incident photon, while the external ones are designed to absorb it after the interaction. The former are usually made of a plastic material, so that scattering becomes dominant over photoelectric absorption above 20~keV and the instrument can reach a lower energy threshold; the latter can either be identical to simplify the design or they can be made of an inorganic crystal to have a higher stopping power and spectral resolution. The line which connects the centers of the two hit bars provides the best approximation of the azimuthal angle $\varphi$, which provides the azimuthal response of the instrument; the histogram of the scattering directions, called modulation curve, shows a cosine square modulation in case of linearly polarized photons and a statistically--flat distribution for unpolarized radiation. The scattering angle $\theta$ can not be measured in this kind of instruments because the depth at which both the scattering and the absorption occur is not known, and in this sense they are two--dimensional.

The design which we propose and are investigating thanks to a dedicated funding provided by the Italian National Institute for Astrophysics (INAF) is compared with the classical one in Figure~\ref{fig:current_vs_proposed}. Scintillator bars are read-out at the two ends: the difference of the light collected by the two sensors provides an estimate of the Depth of Interaction (DoI), as the amount of scintillation light arriving at each ends decreases exponentially with the distance. Assuming a reference axis with center in the center of the bar and direction from the top to the bottom of the bar, and modeling the light attenuation along the bar as an exponential process, we have that the signal collected at each end is:
$$\left\{
\begin{array}{ll}
    A_{\T} = A_0 \exp^{-a(L/2 + x)} \\
    A_{\B} = A_0 \exp^{-a(L/2-x)} 
\end{array}
\right. \quad \longrightarrow \quad E_{\text{deposit}} \propto \sqrt{A_{\T} A_{\B}}  \quad \text{and} \quad x \propto \ln\frac{A_{\B}}{A_{\T}} \label{eq:doi}$$
The knowledge of the depth of interaction for both the scattering and absorption interactions would allow to constrain the scattering angle, provided that the source position, that is, the angle $\delta$ in the figure, is known. In addition to better controlling the systematic effects of the instrument for off-axis sources, this would also be useful to distinguish photons scattered more closely to 90$^\circ$: these are the events which have the larger response to polarization, and weighting them more in the analysis may allow for a significant increase of the sensitivity. In case of photoelectric polarimeters, an appropriate weighting was proved to increase the sensitivity of $\sim$10\%\cite{DiMarco2022}.

Depth of interaction determination is a standard procedure (see, e.g., Chapter 10 in \citenum{Knoll2000}), but it is usually applied when the amount of scintillation light generated is large: incidentally, a better spatial resolution is achieved by decreasing the attenuation length in a controlled way. For the application in hard X-ray polarimetry, it is imperative to achieve a threshold for the detection of scattered photons as low as possible, that is, a few tens of keV, as the flux from astrophysical sources decreases very quickly with energy, typically as a power law of index $\sim$2. Then, to favour Compton scattering against the competing photoabsorption, low-atomic-number plastic scintillators are preferable to inorganic ones as the scattering medium, but the former have a significantly lower light yield. Moreover, the energy deposits for scattering of photons with energy of a few tens of keV is very low: for example, it is of $<$2~keV for 90~$^\circ$ scattering of 30~keV photons, which corresponds to the generation of $<$20~optical photons for the typical light yield of plastic scintillators ($\sim10^4$ ph/MeV). The tiny amount of optical photons typically available in these conditions makes challenging to apply DoI measurement and then we are testing experimentally to what extent this methodology can be applied in the field of hard X-ray polarimeters. 

\section{TEST SET-UP}

\begin{figure} [ht]
   \begin{center}
   \begin{tabular}{ccc} 
   \includegraphics[totalheight=4.5cm]{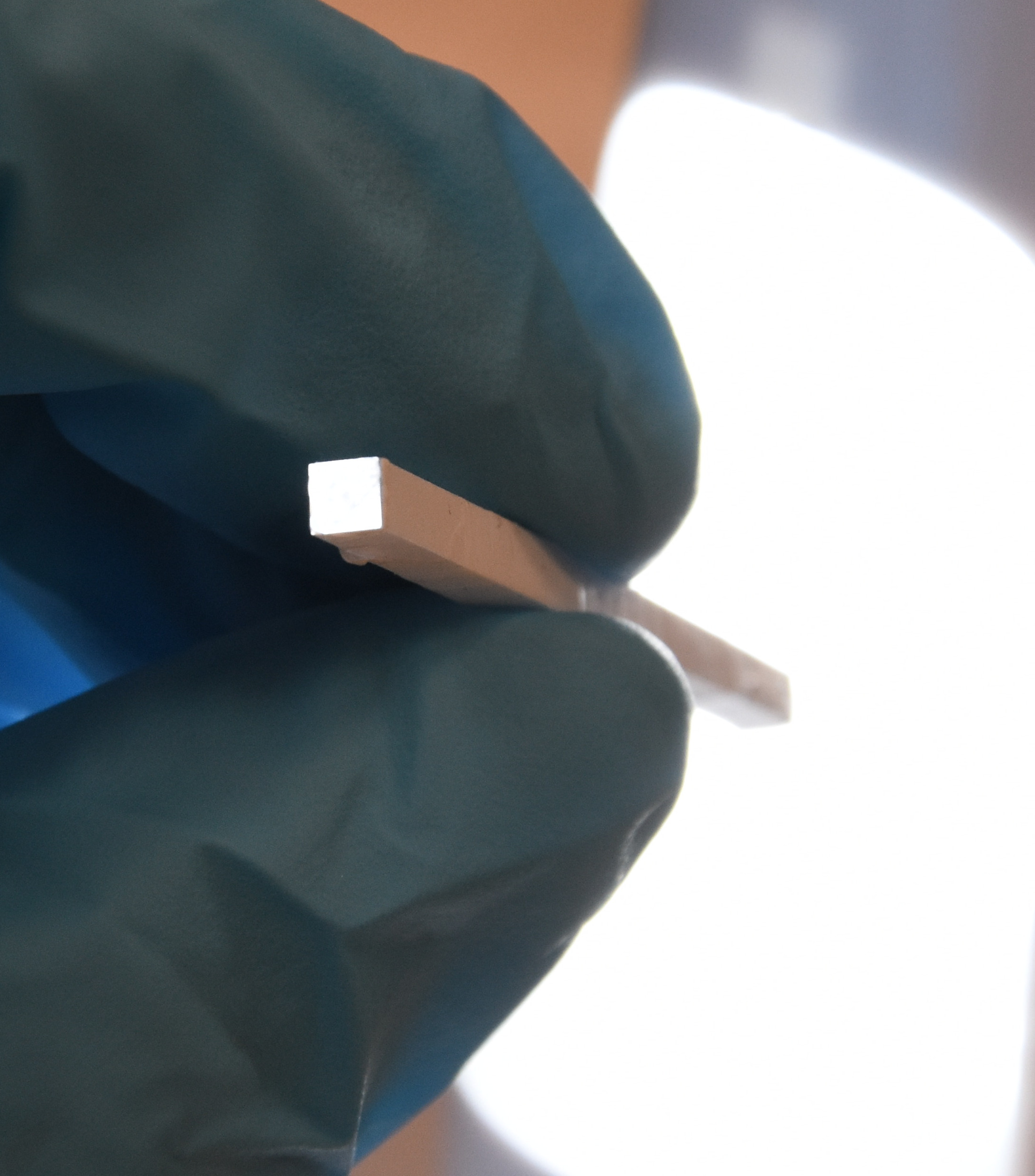} 
   \includegraphics[totalheight=4.5cm]{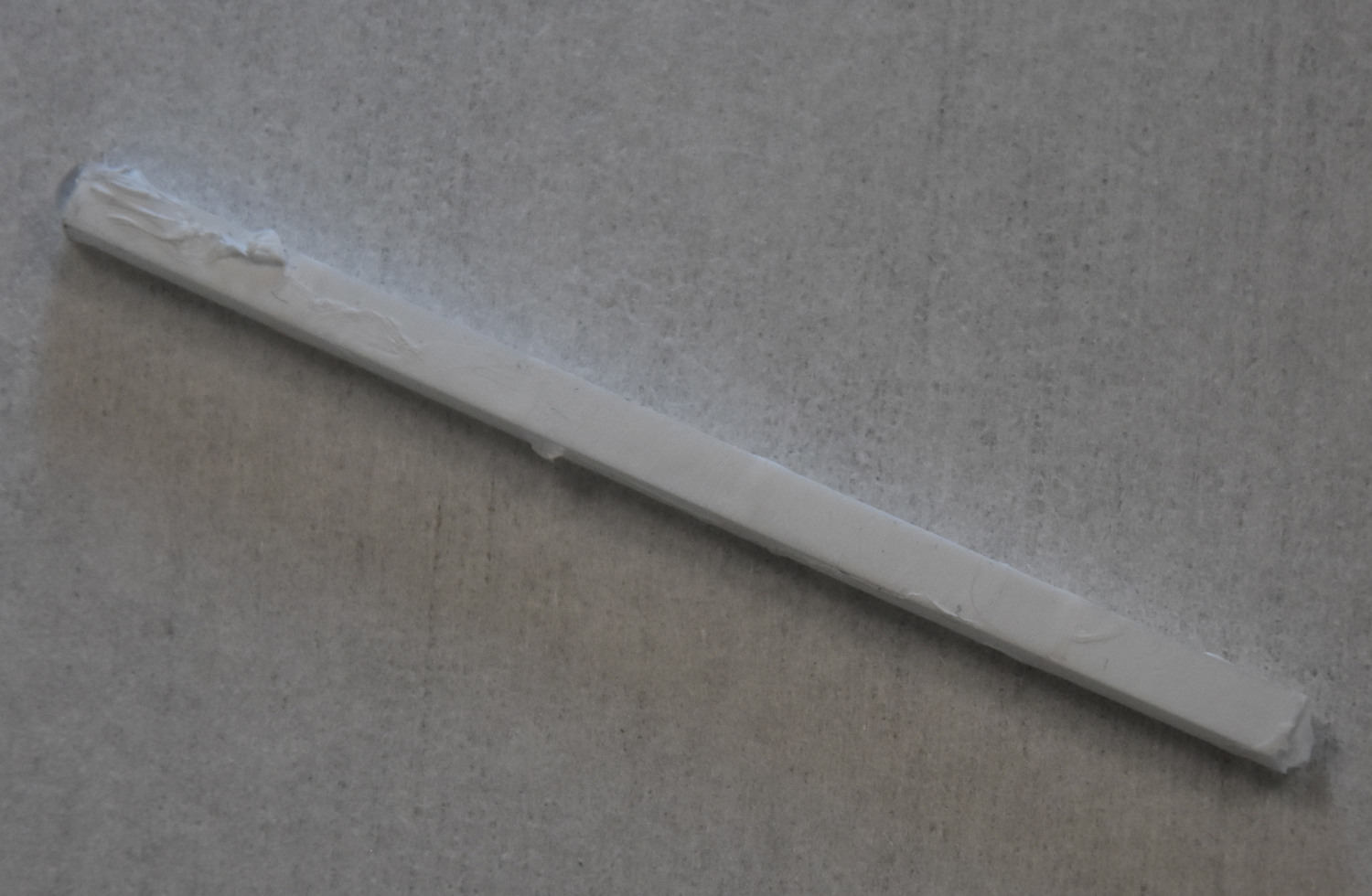}
	\end{tabular}
	\end{center}
   \caption{Plastic scintillator bar made of EJ-204, with size $3\times3\times60~\text{mm}^2$. The bar is wrapped in teflon and optically coupled with grease. \label{fig:scintillator_bar}}
   \end{figure}

We are investigating two different technologies for the double-ends read-out of plastic scintillator bars, i.e., Silicon Drift Detectors\cite{Gatti1984, Vacchi2022} (SDDs) and commercial Silicon Photomultipliers (SiPMs) from Hamamatsu. The goal for both is to test the read-out of a scintillator bar with size $3\times3\times60~\text{mm}^2$ made of EJ-204, which is a general-purpose plastic scintillator material producing 10,400 optical photons per MeV (see Figure~\ref{fig:scintillator_bar}). The wavelength of maximum emission is 408~nm.

\subsection{Silicon Drift Detectors}

Arrays of SDDs detectors are being developed in Italy by the INFN/FBK-led ReDSoX collaboration\footnote{\url{http://redsox.iasfbo.inaf.it/redsox/}}, with a large INAF contribution. Initially developed for application as particle trackers\cite{Gatti1984, Vacchi1991, Rashevsky2002}, SDDs were later proved to be sensitive also to X-rays\cite{Zampa2011,Campana2011} and then to scintillation light, with the capability of detect simultaneously X-rays with a proper signal discrimination\cite{Fuschino2020}. SDDs use low-noise read-out electronics to achieve an energy threshold lower than 1~keV for X-rays, also in case the sensor has a very large area of tens of cm$^2$. They have been qualified for space use, e.g. onboard the first payload of the HERMES constellation\cite{Fiore2020}, launched with the SpIRIT mission\cite{Thomas2023}. Moreover, they are expected to be used on-board the THESEUS mission, a candidate for the 7th medium size ESA Cosmic Vision call\cite{Amati2021}.

\begin{figure} [ht]
   \begin{center}
   \subfloat[\label{fig:SDDs}]{\includegraphics[totalheight=4.3cm]{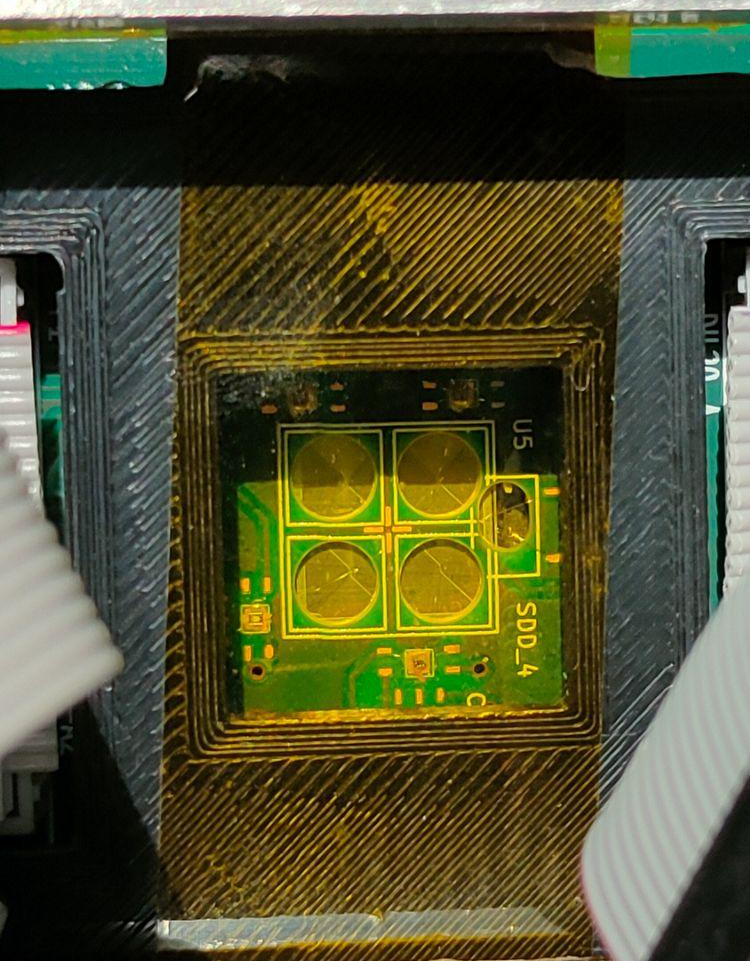}} \hspace{1mm}
   \subfloat[\label{fig:SDDs_setup}]{\includegraphics[totalheight=4.3cm]{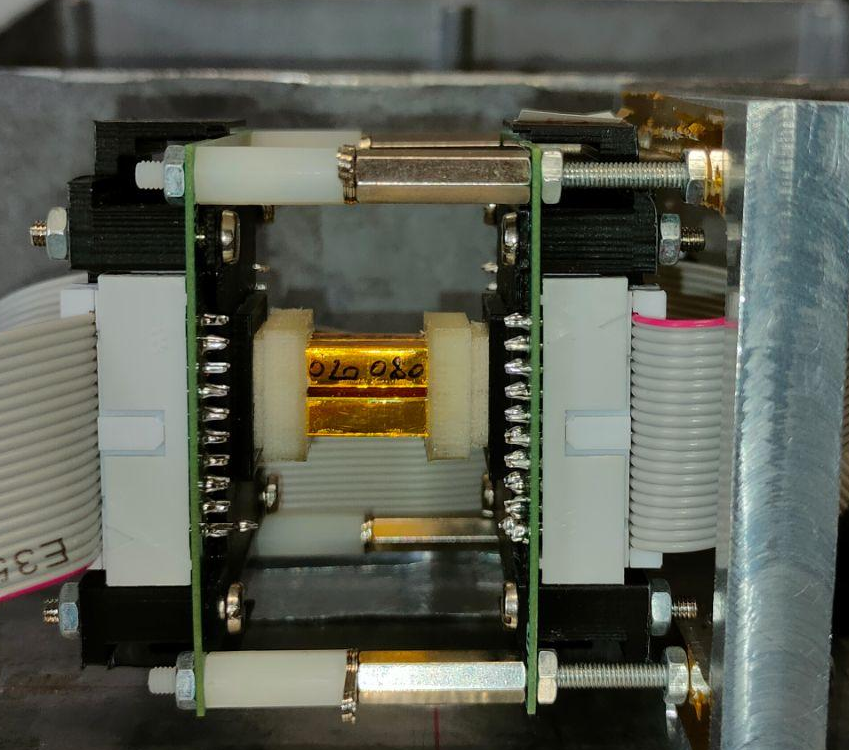}} \hspace{1mm}
   \subfloat[\label{fig:SDDs_table}]{\includegraphics[totalheight=4.3cm]{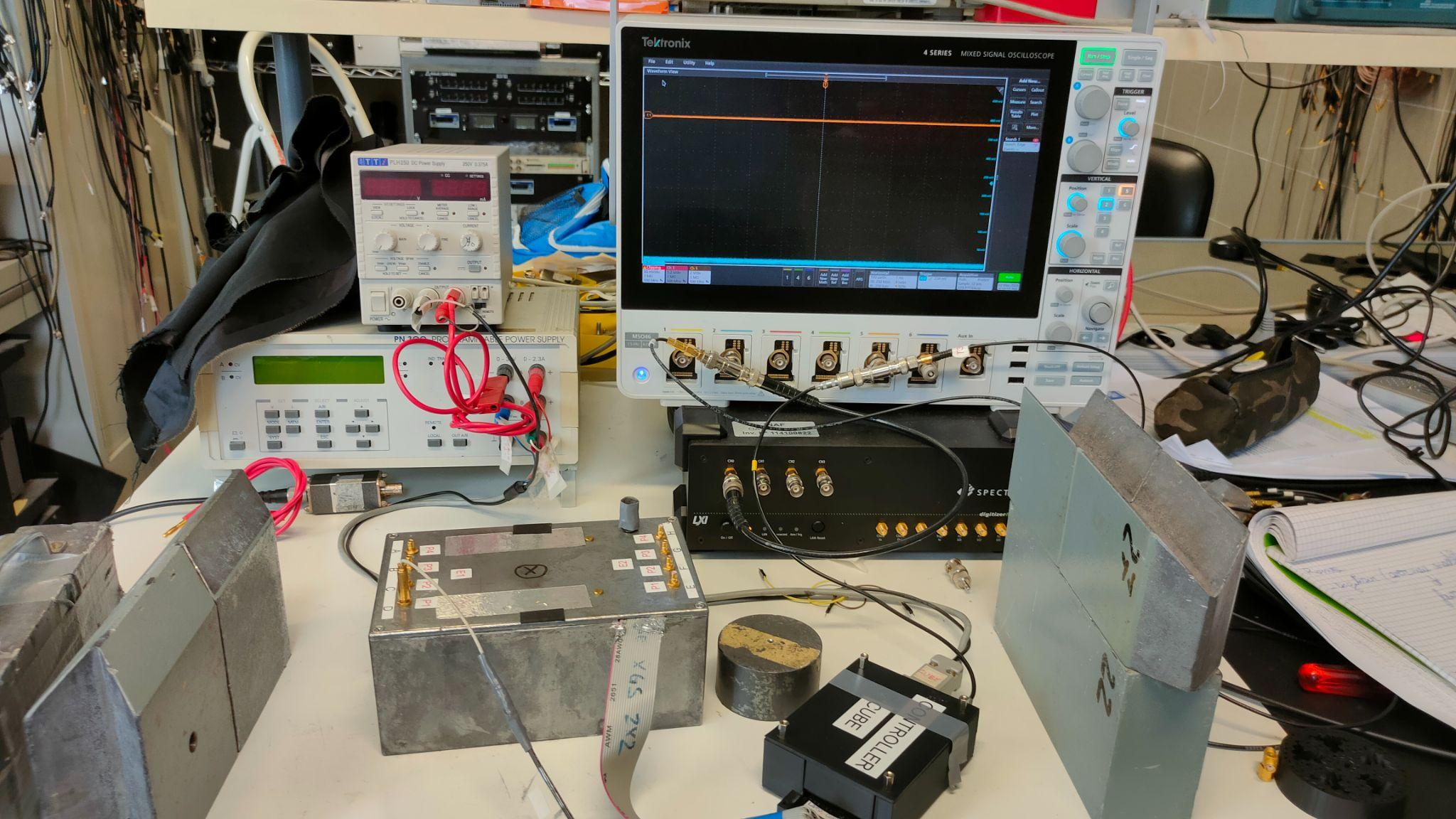}}
   \end{center}
   \caption{Overview of the set-up based on SDDs. (a) One of the two arrays of $2\times2$ SDDs (with size $5\times5~\text{mm}^2$ each) used to read-out up to four scintillator bars. (b) Two of such arrays, with 4 scintillator bars in-between, form the double-end read-out. (c) The set-up on the test bench at INAF/OAS in Bologna.}
   \end{figure}

Our test set-up is based on 2 arrays of $2\times2$ SDDs with size $5\times5~\text{mm}^2$ each (see Figure~\ref{fig:SDDs}), read-out by CUBE preamplifiers. The two arrays are coupled to 4 scintillators bars at the two ends (see Figure~\ref{fig:SDDs_setup}), which currently are made of CsI(Tl) with size $5\times5\times30~\text{mm}^2$ for testing purposes. The signal is digitized at 16 bits by a Spectrum Instrumentation digitizer (DN2.442-04, capable of sampling rates up to 250 MS/s). This setup is currently being  programmed to read a customizable combination of the input signals when they are above a selectable threshold within a certain time window, and save the filtered signals to file in the HDF5 format. The overall set-up at INAF/OAS is shown in Figure~\ref{fig:SDDs_table}.

\subsection{Silicon Photomultipliers}

Our second set-up is based on commercial Silicon Photomultiplier by Hamamatsu\cite{DelMonte2016}. Two Multi-Pixel Photon Counters (MPPCs, which is the commercial name of SiPMs sold by Hamamatsu) model S13360-3050CS are read-out by a control board model C12332, specifically developed by Hamamatsu for the evaluation of these sensors. The control board allows for the compensation of the gain due to temperature variations and it has a built-in amplifier with nominal gain of 100; the characteristics of the sensors are reported in Table~\ref{tab:mppc}. The devices are operated with a low overvoltage, $\lesssim2$~V, to reduce the dark current.

    \begin{table}[ht]
    \caption{Main characteristics of the tested MPPC/Si-PM by Hamamatsu.} 
    \label{tab:mppc}
    \begin{center}       
    \begin{tabular}{|c|c|}
    \hline
    Effective area & $3\times3~\text{mm}^2$ \\ \hline
    Number of pixels & 3600 \\\hline
    Pixel size & 50~\textmu m \\\hline
    Spectral response range & 270 to 900 nm \\\hline
    Peak sensitivity wavelength (typ.) & 450 nm \\\hline
    Dark count/ch (typ.) & 500 kcps \\\hline
    Fill factor & 74\% \\\hline
    Window & Silicon epoxy, n=1.41 \\\hline
    \end{tabular}
    \end{center}
    \end{table}
    
\begin{figure} [ht]
   \begin{center}
   \begin{tabular}{cc} 
   \includegraphics[totalheight=7.5cm]{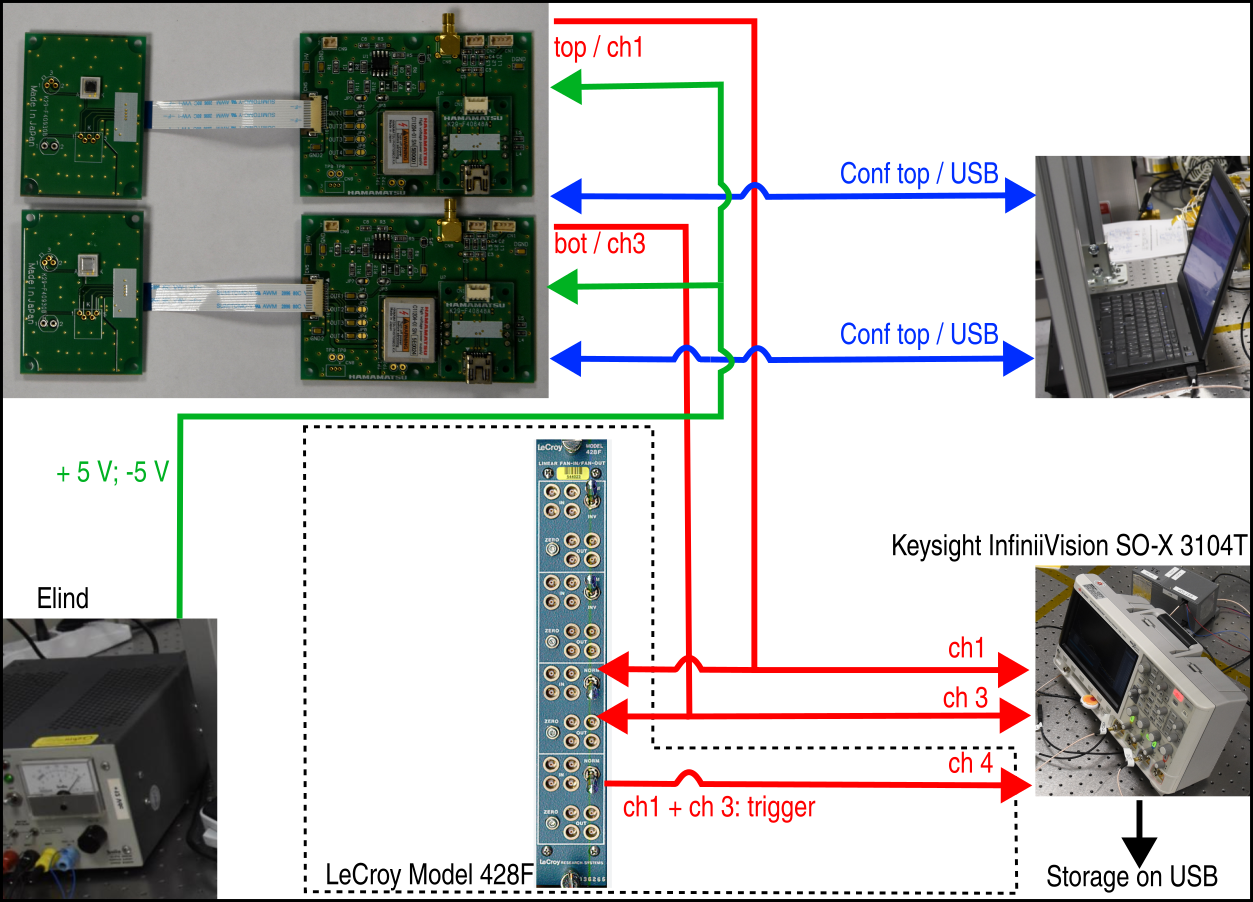}
	\end{tabular}
	\end{center}
   \caption 
   {Block diagram of the SiPM set-up. The two detectors read-out the light signal from the top and bottom end of the scintillator bar. The analog signal is digitized with a oscilloscope, and the acquisition is triggered when the \emph{sum} of the bottom and top signals is larger than the chosen threshold. \label{fig:sipm_block_diagram}}
   \end{figure}

\begin{figure} [ht]
   \begin{center}
   \begin{tabular}{cc} 
   \subfloat[\label{fig:DSC_6336}]{\includegraphics[angle=90,totalheight=6.5cm]{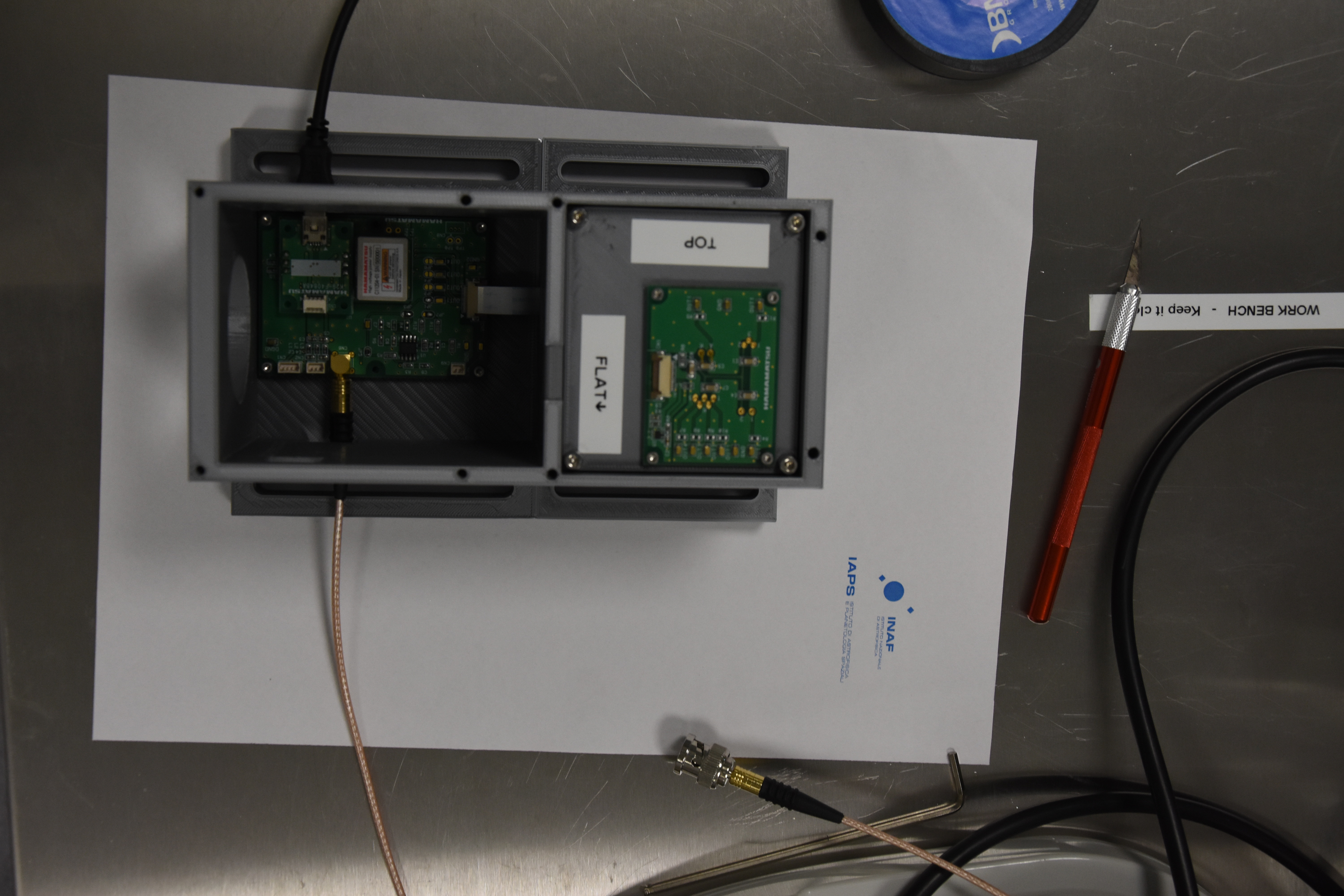}} \hspace{2mm}
   \subfloat[\label{fig:DSC_6398}]{\includegraphics[angle=0,totalheight=6.5cm]{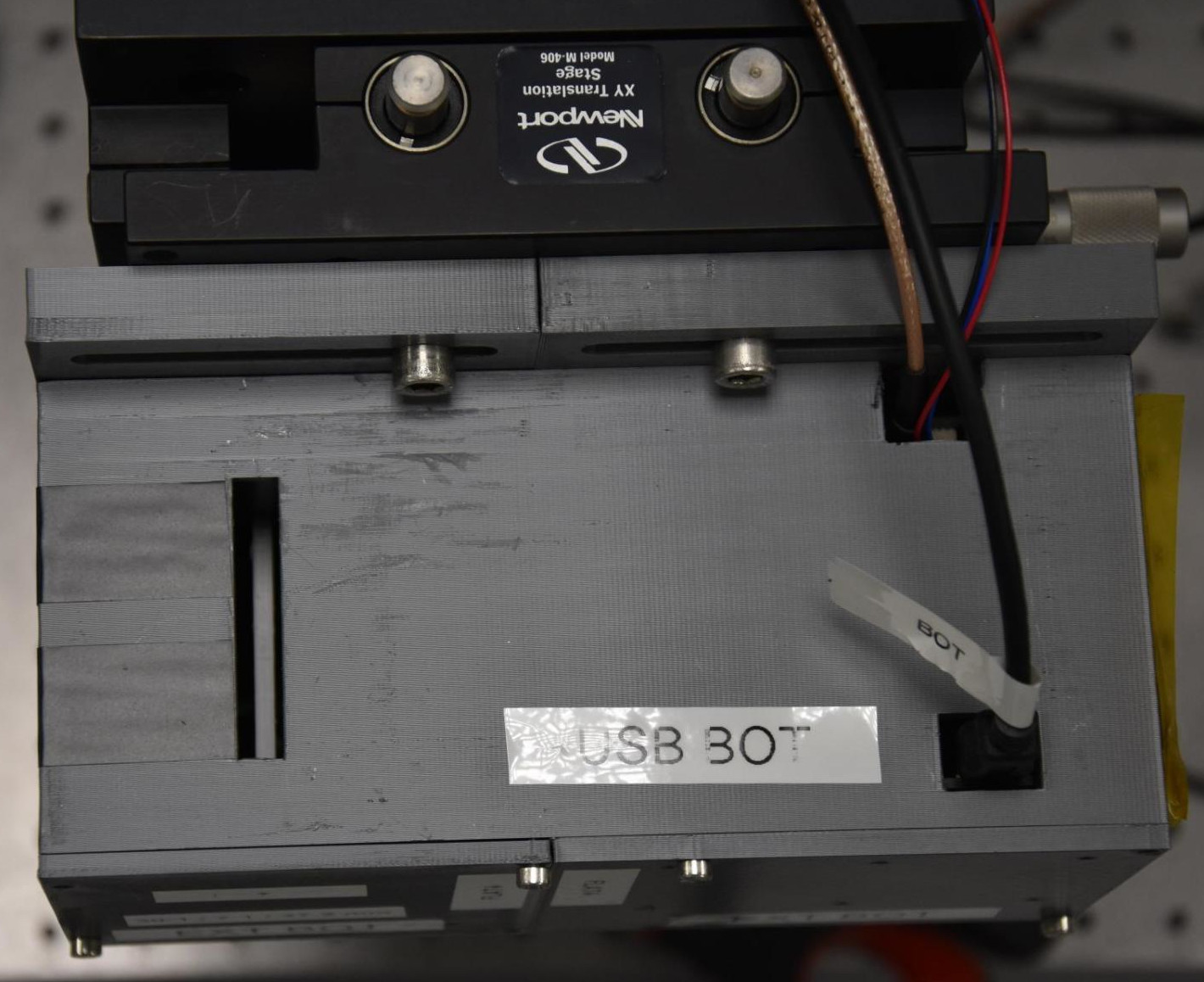}}
	\end{tabular}
	\end{center}
   \caption{(a) Dark housing box for the SiPMs during mounting. The bottom control board and the (rear) of the board hosting the top SiPM are visible. (b) Assembled dark box with a view on the plastic scintillator bar before mounting the optically opaque X-ray window. \label{fig:sipm_box}}
   \end{figure}

A block diagram of the set-up is shown in Figure~\ref{fig:sipm_block_diagram}. Two SiPMs are used to read-out the top and bottom end of the plastic scintillator bar. They are connected to a single power supply: the bias voltage applied to each SiPM, generated by the control board, can be set through an USB connection to a laptop computer. The SiPM analog signal is fed into an oscilloscope (Keysight InfiniiVision SO-X 3104T), which is essentially used as a digitizer with a sampling rate up to 2.5 GS/s. The acquisition of both the top and bottom signals is triggered when the \emph{sum} of the two (which is done analogically by a Fan-In/Fan-Out NIM LeCroy 428F module), is larger than the chosen threshold. The two SiPMs and the scintillator bar are mechanically mounted in a dark box (see Figure~\ref{fig:sipm_box}), with an entrance window for X-rays made of three layers of 12.5~\textmu m kapton coated with 100~nm of aluminum.

\begin{figure} [ht]
   \begin{center}
   \begin{tabular}{cc} 
   \subfloat[\label{fig:cross}]{\includegraphics[totalheight=5.2cm]{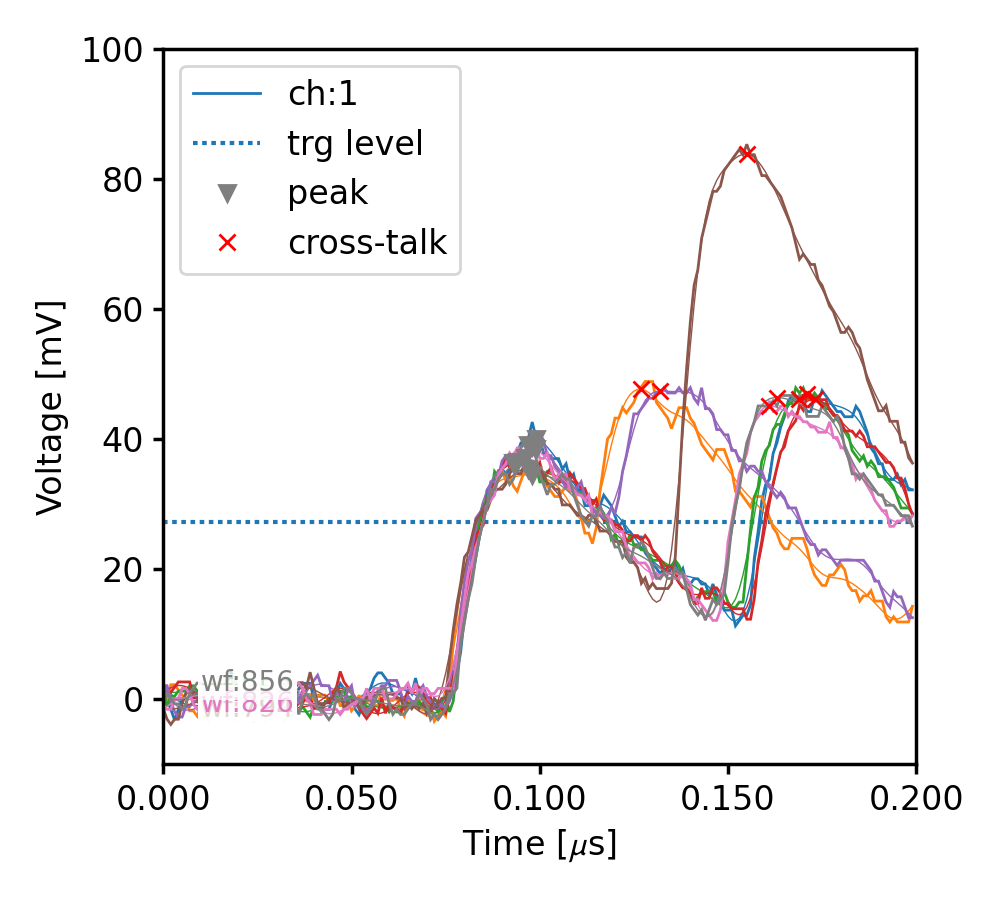}}
   \subfloat[\label{fig:peak_spectrum}]{\includegraphics[totalheight=5.2cm]{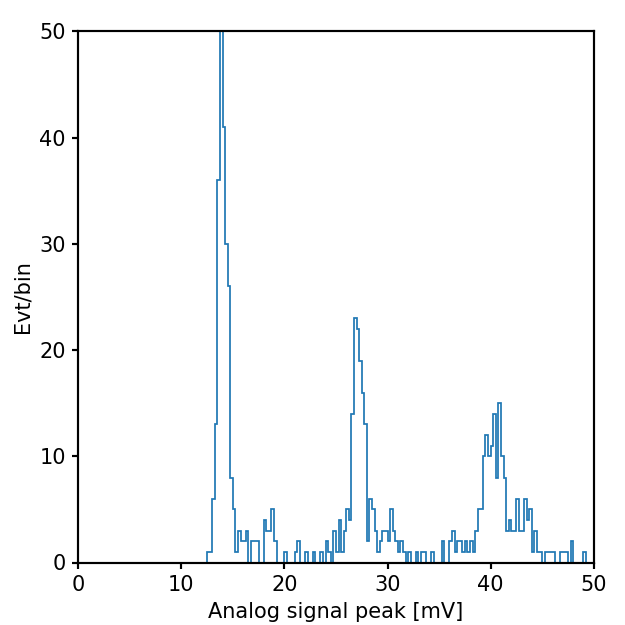}}
   \subfloat[\label{fig:coincidence}]{\includegraphics[totalheight=5.2cm]{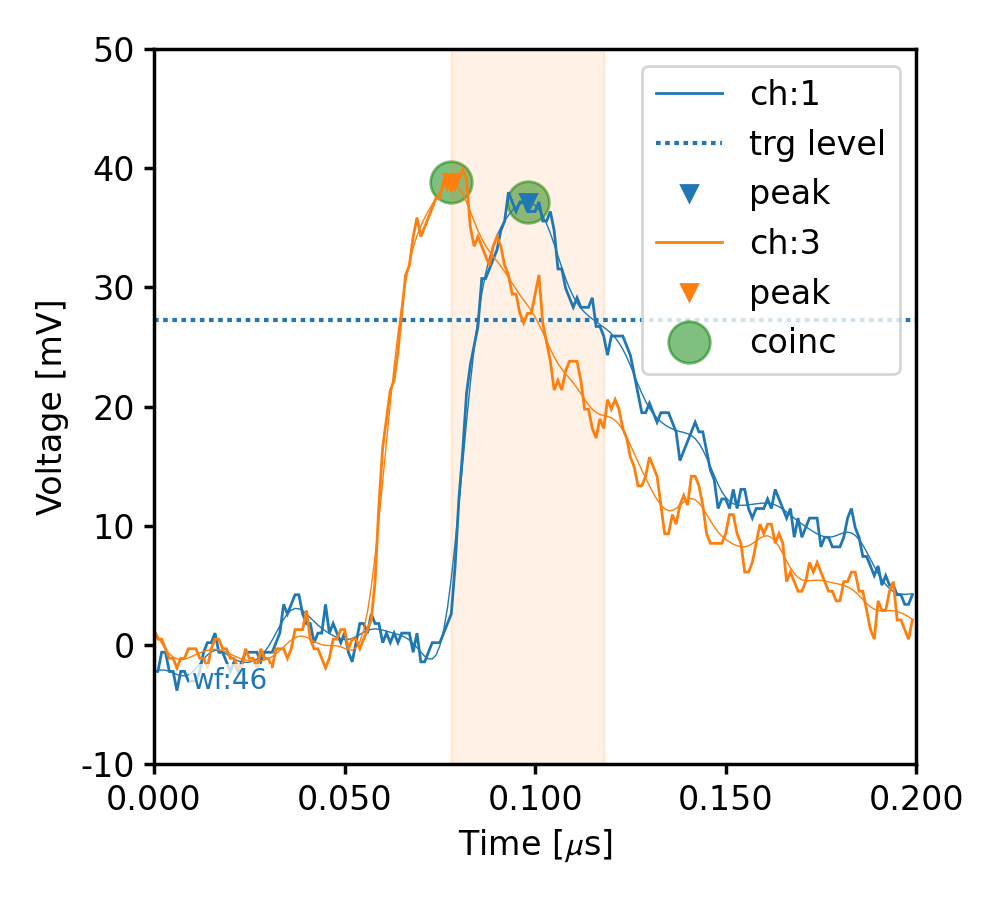}}
   
	\end{tabular}
	\end{center}
   \caption{(a) Examples of waveforms digitized with the oscilloscope. Only the first peak is considered: the subsequent ones are labelled and neglected in the analysis. (b) Histogram of the identified peaks, which are fitted to derive the gain per photoelectron and convert amplitudes from volts to the number of photoelectrons. (c) The last step of the analysis is to select those events which produce a signal in both the top and bottom channels within a certain time window.}
   \end{figure}

Digitized signals, or waveforms, are processed off-line. After the application of a low-pass filter, the waveform of both top and bottom channels are analyzed to search for peaks. In case more than one peaks is found, which occurs in $\lesssim5\%$ of the cases, only the first is considered for further analyses (see Figure~\ref{fig:cross}). For each channel, the histogram of the peak amplitudes is built to derive the gain between the signal amplitude and the number of emitted photoelectrons (see Figure~\ref{fig:peak_spectrum}). Such a value is used to convert the peak amplitude from voltage to the number of emitted photoelectrons, or p.e.. The last step is look for those signals which occur in both bottom and top channels within a certain time window, which is typically of $\pm$10~ns and it is conventionally centered at the time of the peak in the top channel.

\section{Preliminary results}

\begin{figure} [ht]
   \begin{center}
   \begin{tabular}{cc} 
   \subfloat[\label{fig:box_on_ice}]{\includegraphics[totalheight=5.3cm]{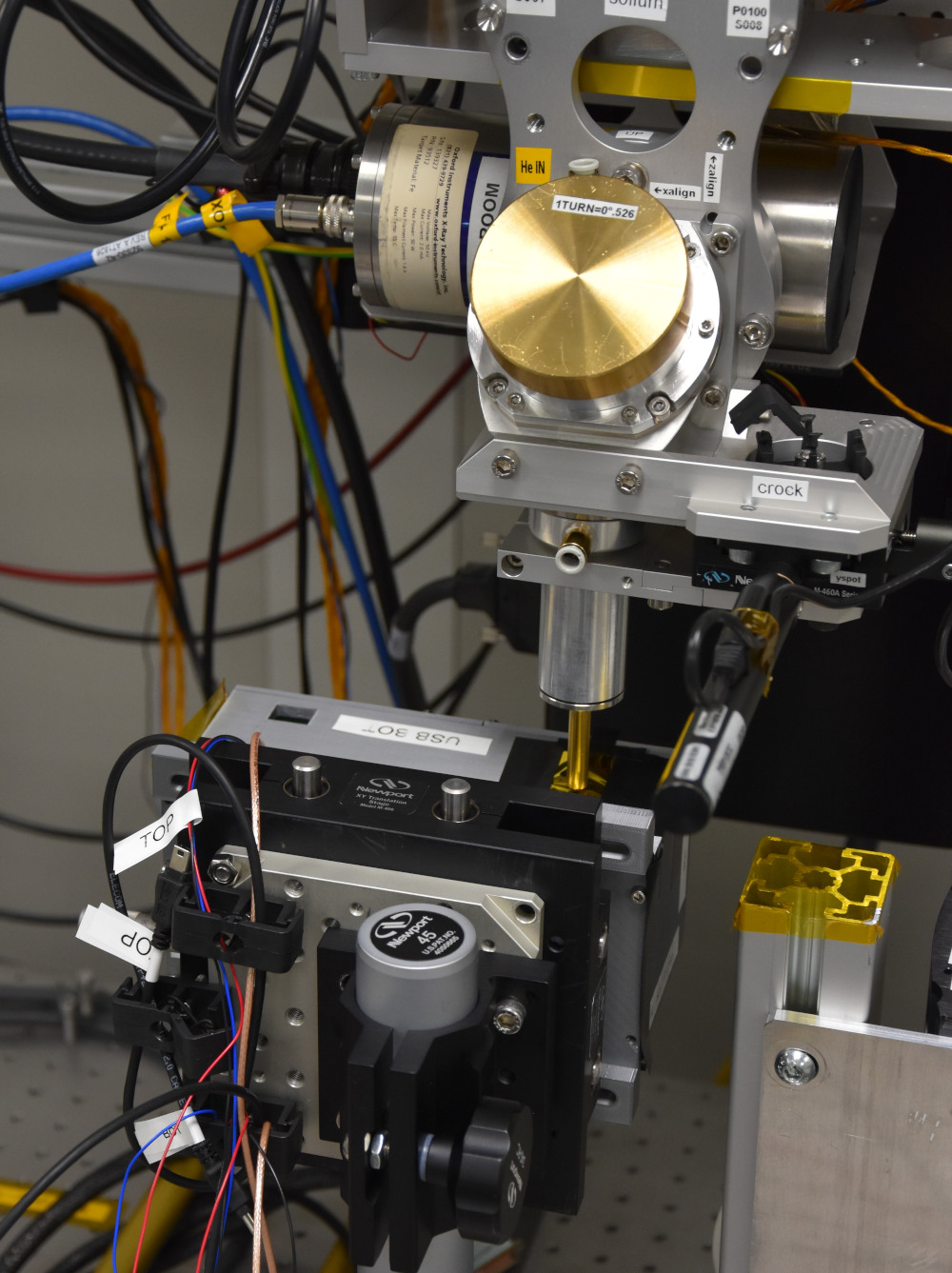}} \hspace{2mm}
   \subfloat[\label{fig:spot_6p4keV}]{\includegraphics[totalheight=5.3cm]{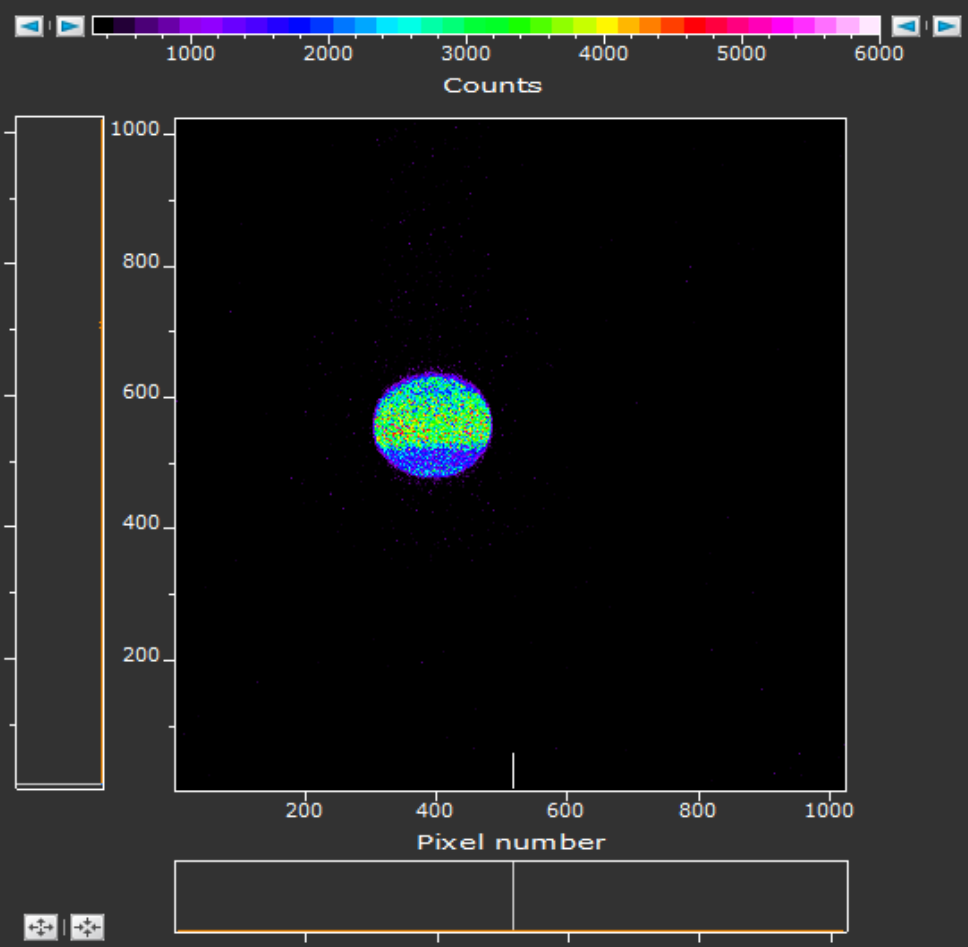}}\hspace{2mm}
   \subfloat[\label{fig:spectrum_6p4keV}]{\includegraphics[totalheight=5.3cm]{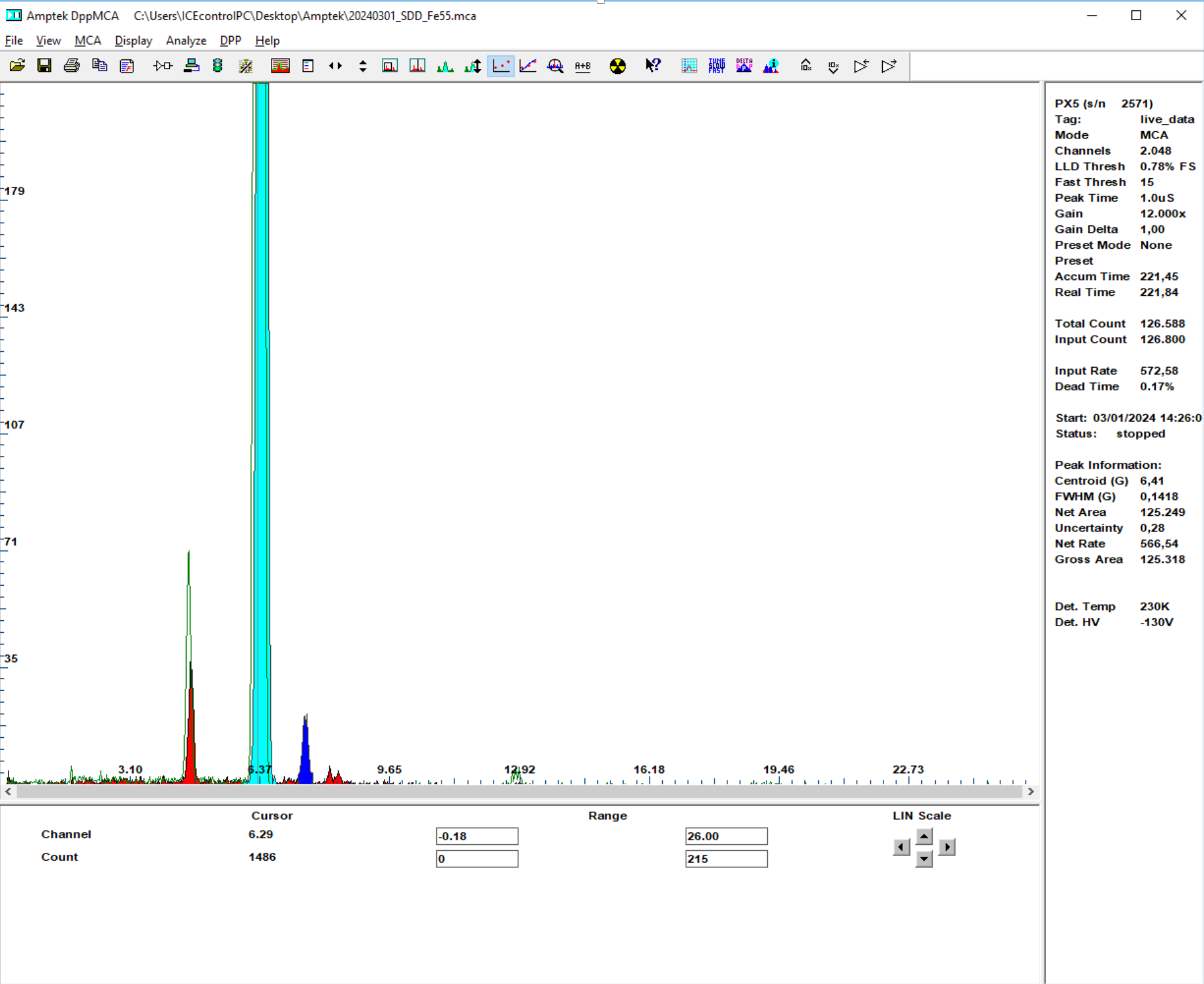}}
	\end{tabular}
	\end{center}
   \caption{(a) Dark box containing the SiPM set-up mounted on the ICE. The source on the top is emitting nearly monochromatic X-rays at 6.4~keV by diffracting the emission from a tube with iron anode on a Si (400) crystal. (b) Image of the source spot acquired with an Andor iKon-M~SY CCD. The spot is not uniform because of the Bragg diffraction on the Si (400) crystal. (c) Source spectrum measured with an Amptek FAST SDD with size $7\times7~\text{mm}^2$. The main line is that at 6.4~keV; the small line on the left of it is the escape peak from the silicon detector, while the one at higher energy corresponds to the fluorescence from the nichel contained in the brass collimator.}
   \end{figure}

The SiPM setup is being tested with X-rays in the Instrument Calibration Equipment (ICE) available at INAF-IAPS\cite{Muleri2022b}. The scintillator bar is illuminated with nearly monochromatic X-rays at 6.4~keV, which are photoabsorbed in the scintillation material, thus producing a charge deposit of this energy (see Figure~\ref{fig:box_on_ice}). Incident X-rays are generated by diffraction of X-ray emission from a tube with Fe anode on a Si (400) crystal, setting the HV of the X-ray tube to 11 keV. The spot is circular with $\sim$2~mm diameter (see Figure~\ref{fig:spot_6p4keV}), and the incident spectrum comprise of 6.4~keV photons for more than 99\% (see Figure~\ref{fig:spectrum_6p4keV}). The spot is placed at the center of the scintillator bar, which is 60~mm long.

\begin{figure} [ht]
   \begin{center}
   \begin{tabular}{cc} 
      \subfloat[\label{fig:coinc_rate}]{\includegraphics[totalheight=5.cm]{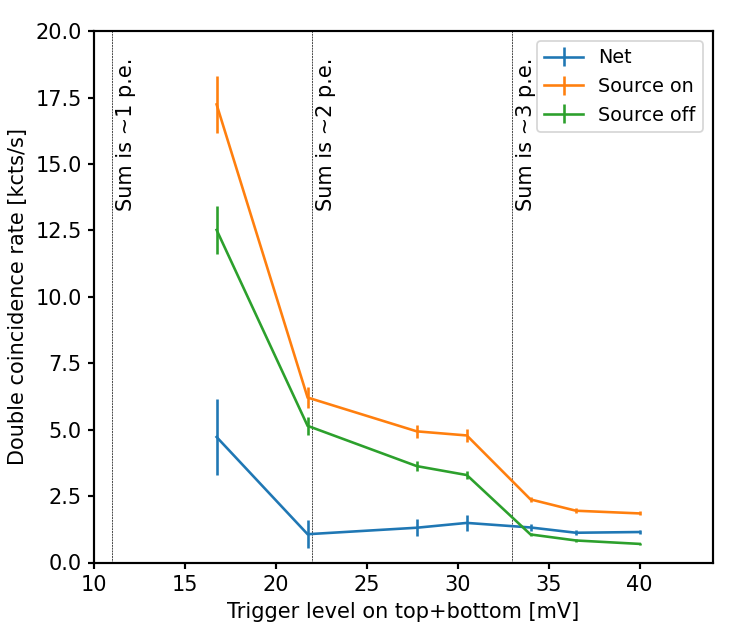}}\hspace{2mm}
   \subfloat[\label{fig:coinc_spectrum}]{\includegraphics[totalheight=5.cm]{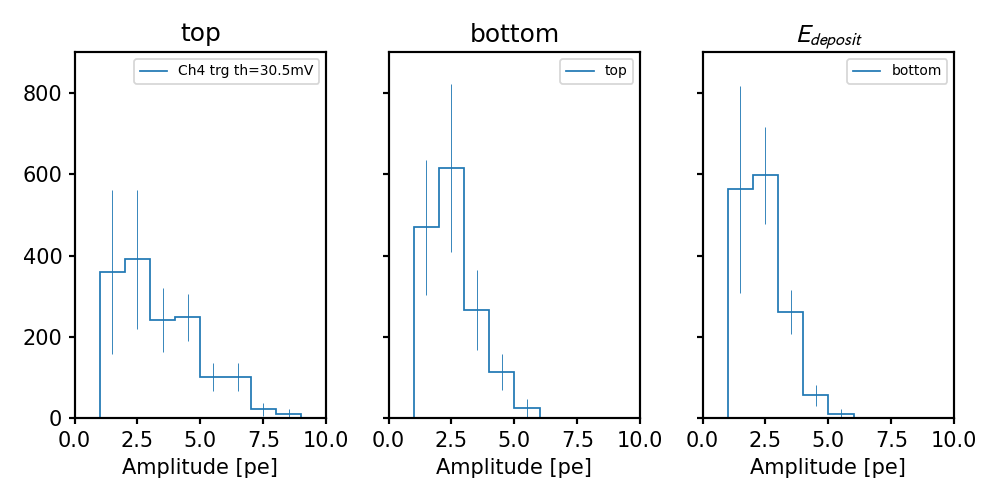}}
	\end{tabular}
	\end{center}
   \caption{(a) Rate of coincidences within a time window of $\pm$10~ns when the source is off (green line) or on (orange line). The difference between the two curves (``net'' coincidences) is in blue. (b) Source spectrum, obtained by subtracting that of the spurious coincidences measured when the source is turned off, measured for the bottom and top channels, plus the estimate of the energy deposit with Equation~\ref{eq:doi}.}
   \end{figure}

We show in Figure~\ref{fig:coinc_rate} the rate of coincidences between the top and bottom channels as a function of the triggering threshold on the sum of the bottom and top signals. The rate of spurious coincidences $r_c$ when the threshold is larger than 2 photoelectrons, which corresponds to a signal of at least of 1 p.e. for each of the two channels, is $<10^4~\text{evt}/s$. This is in line with the expectations derived by the nominal dark current of the single SiPMs ($\sim5\times10^5~\text{evt}/s$, see Table~\ref{tab:mppc}) and the 20~ns time windows, which is (see Chapt.~17 of \citenum{Knoll2000}):
\begin{equation}
    r_c^{\text{exp}} = 2 r_1 r_2 \delta t = 10^4 \text{\,\,evt/s}
\end{equation}
The rate of spurious coincidences decreases to $2\times10^3$~evt/s when the threshold is increased to 3 p.e., corresponding to a signal of 1 p.e. in one channel and of 2 p.e. in the other. The source rate of $\sim10^3$~evt/s is consistently detected for different thresholds when the rate of (spurious) coincidences with the source switched off is subtracted to the rate obtained when the source is on (``net'' data points in Figure~\ref{fig:coinc_rate}).

We also show in Figure~\ref{fig:coinc_spectrum} the source spectrum, after subtracting the spurious coincidences obtained when the source is turned off, measured for the bottom and top channels, plus the estimate of the energy deposit with Equation~\ref{eq:doi}. Energy deposits of 6.4~keV generate a signal of up to several photoelectrons, with an average of just a few of them at each end of the scintillator bar.
   
\section{Conclusions}

We have presented the preliminary status of our activities for the evaluation of a double-end set-up for the read-out of plastic scintillator bars. This technique is already routinely used when there is a large energy deposit in the scintillating material to determine the Depth of the Interaction, but our aim is to test the low-energy applicability of this technique for its use in Compton polarimeters dedicated to the observation of cosmic sources. We selected two technologies for reading the scintillation light: Silicon Drift Detectors, developed in Italy by a large collaboration for the measurement of also X-rays and charged particles, and commercial Silicon photomultipliers. While we are refining the read-out chain of the former, we used off-the-shelves electronics to test the latter and we presented the first results obtained. We read-out a plastic scintillator bar with size $3\times3\times60~\text{mm}^2$ made of EJ-204 at the two ends with two $3\times3~\text{mm}^2$ MPPCs from Hamamatsu. An energy deposit of 6.4~keV, corresponding to $90^\circ$ scattering of $\sim60$~keV photons, is generated at the center of the bar by photo-absorbing X-rays from a monochromatic source based on Bragg diffraction. Such a deposit is able to generate in both the MPPCs a signal of a few photoelectrons, which is discernible once only the signals occurring in time coincidence (within $\pm$10~ns) are selected. While spurious coincidences are still present in the tested configuration and would cause an undesirable background in a real implementation of a Compton polarimeter, in such a configuration a third time coincidence with the signal in the absorber will be possible. The capability to suppress the background in this case will be the subject of future developments of our work. 

\acknowledgments 
 
Acknowledgement: This activity is financially supported by INAF though a Large Grant with title ``The Compton polarimeter on board the Soft-X and gamma-ray WIde-field Polarimeter Explorer'' in the context of the ``Bando di Astrofisica Fondamentale 2022'' program 

\bibliography{References.bib} 
\bibliographystyle{spiebib} 

\end{document}